\documentclass[prl,twocolumn]{revtex4}

\usepackage{amssymb}
\usepackage{latexsym}
\usepackage{amsmath}

\newcommand{\bea}{\begin{eqnarray}}
\newcommand{\eea}{\end{eqnarray}}

\begin{document}
\onecolumngrid{
\hfill\parbox{2.8cm}
{DAMTP-2005-19 \\  hep-th/0502124}}\\

\title{Integrability and the Kerr-(A)dS black hole in five dimensions}
\author{Hari K. Kunduri} \email{H.K.Kunduri@damtp.cam.ac.uk}
\author{James Lucietti} \email{J.Lucietti@damtp.cam.ac.uk}
\affiliation{
DAMTP, Centre for Mathematical Sciences, \\
University of Cambridge,
Wilberforce Road,
Cambridge CB3 0WA, UK}

\begin{abstract}
In this note we prove that the Hamilton-Jacobi equation for a
particle in the five dimensional Kerr-(A)dS black hole is separable,
for arbitrary rotation parameters. As a result we find an irreducible Killing
tensor.  We also consider the Klein-Gordon
equation in this background and show that this is also
separable. Finally we comment on extensions and implications of these results.
\end{abstract}

\maketitle

\section{Introduction}
It is a remarkable fact that the geodesic equation of the four
dimensional Kerr black hole is integrable, in the sense of Liouville~\cite{Carter}. The result is at first unexpected. The black hole has
two Killing vectors which, together with the geodesic constraint,
give three constants of motion in the four dimensional configuration space
of a particle moving in this background.  Nevertheless one can show
there exists an extra conserved quantity rendering the system
integrable. This `hidden symmetry' is due to another constant of motion
on the phase space of the particle and furthermore may be thought of
as resulting from non trivial properties of the space-time.  These
results were generalised to include the case of a rotating black hole
with non-zero cosmological constant~\cite{Carter2}.

In this paper we will consider the integrability properties of two
important partial differential equations defined on the space-time
consisting of the five dimensional Kerr-(A)dS black hole with two
arbitrary angular momentum parameters. Namely, we will consider the
Hamilton-Jacobi (HJ) equation and the Klein-Gordon (KG) equation.
The first is of relevance in classical mechanics, and its
separability implies that geodesic motion on the space-time is
integrable. The KG equation is of course relevant when considering
quantum theory on the space-time. The separability of these
equations is closely related to geometric properties of the
manifold, in particular the existence of second rank Killing
tensors. This is a symmetric tensor $K_{\mu\nu}$ which satisfies
$\nabla_{(\mu} K_{\nu\rho)}=0$. Note that unlike Killing vectors,
Killing tensors do not give rise to Noether charges of Lagrangian
theories built on the space-time with the corresponding metric. They
are symmetries which can only be seen in phase space.

Our work generalises previous work
as follows. Firstly, the case where the angular momenta are equal
can be read off from the results of~\cite{Page}, in which these two
equations were shown to be separable in all odd dimensional
Kerr-(A)dS spacetimes when all rotation parameters are equal. The
Killing tensor found however turned out to be \emph{reducible} in
the sense it can be written as a linear combination of direct
products of Killing vectors. In this degenerate case, the enhancement
of symmetry of the space-time is enough to ensure that the HJ equation
separates, thus rendering this result less surprising.

Secondly, a non trivial Killing tensor was shown to exist in the
case with only one non vanishing parameter in all
dimensions~\cite{GLP2}. It is clearly a remarkable property of five
dimensions that separability can in fact be proven in the general case.
However, it is not wholly unexpected as the five dimensional vacuum
Myers-Perry black hole has been shown to possess a Killing tensor
for arbitrary angular momentum parameters~\cite{Frolov1,Frolov2}.

The five dimensional Kerr-(A)dS metric was first constructed in~\cite{HHT}
and subsequently generalised to all dimensions in the work
of~\cite{GLP}. The metric is most compactly written in Kerr-Schild
coordinates. However, for our purposes Boyer-Lindquist coordinates
are more suitable due to the absence of any off-diagonal components
involving $dr$. Explicitly the metric is:
\bea ds^2 &=&
- W\, (1 - \lambda r^2)\,
d\tau^2 + \frac{\rho^2\, dr^2}{V-2M} + \frac{\rho^2}{\Delta_{\theta}}d\theta^2 \nonumber \\
&& +\frac{2M}{\rho^2}\left( d\tau - \sum_{i=1}^{2} \frac{a_i\, \mu_i^2\,
d\varphi_i}{
1 + \lambda\, a_i^2} \right)^2 \nonumber \\
&&+ \sum_{i=1}^2 \frac{r^2 + a_i^2}{1 + \lambda\, a_i^2}\,
\mu_i^2\, (d\varphi_i-\lambda\, a_i\, d\tau)^2,\label{BL}
\eea
where we have the definitions:
\bea
&&\rho^2 = r^2+a^2\cos^2\theta +b^2\sin^2\theta, \\
&&\Delta_{\theta} = 1+\lambda a^2 \cos^2\theta +\lambda
b^2\sin^2\theta, \\
&&V =  r^{-2}(1-\lambda r^2) (r^2 + a^2)(r^2 + b^2), \\
&&W= \sum_{i=1}^2 \frac{\mu_i^2}{1+\lambda a_i^2} =
\frac{\Delta_{\theta}}{\Xi_a \Xi_b}
\eea
and $\Xi_a =1+\lambda a^2$, $\Xi_b=1+\lambda b^2$, $a_1=a$,
$a_2=b$, $\mu_1=\sin\theta$ and $\mu_2 =\cos\theta$. This metric
satisfies $R_{\mu\nu}= 4\lambda g_{\mu\nu}$.
We point out that for general rotation parameters $a$ and $b$, this
metric possesses three commuting Killing vectors and has the isometry
group $\mathbb{R}\times U(1) \times U(1)$.
It will be convenient to write down the inverse of the metric at
this stage as we will need it in several instances in the later
sections. Unfortunately it is rather unsightly:
\begin{eqnarray}\label{ginv}
g^{\tau r}&=&g^{\varphi _i r}=0 \,, \nonumber \\
g^{rr}&=&\frac{V-2M}{\rho^2}\,, \nonumber \\
g^{\tau \tau}&=&Q-\frac{4M^2}{\rho^2(1-\lambda r^2)^2(V-2M)}\,, \nonumber \\
g^{\tau \varphi _i}&=&\lambda a_i Q -\frac{4M^2a_i(1+\lambda a_i
^2)}{\rho^2(1-\lambda r^2)^2 (V-2M)(r^2+a_i ^2)} \nonumber \\ &-&\frac{2M}{\rho^2}\frac{a_i}{(1-\lambda
r^2)(r^2+a_i ^2)}\,, \nonumber \\
g^{\varphi _i \varphi_j}&=& \frac{(1+\lambda a_i ^2)}{(r^2+a_i ^2)\mu _i ^2}
\delta ^{ij}+\lambda ^2 a_i a_j Q +Q^{ij}  \nonumber \\
&+&\frac{4M^2 a_i a_j (1+\lambda a_i^2) (1+\lambda a_j
^2)}{\rho^2(1-\lambda r^2)^2(V-2M)(r^2+a_i ^2)(r^2 +a_j ^2)}  \nonumber \\
g^{\theta\theta} &=& \frac{\Delta_{\theta}}{\rho^2}
\end{eqnarray}
where $Q$ and $Q^{ij}$ are defined to be \bea
Q&=&-\frac{1}{W(1-\lambda r^2)}-\frac{2M}{\rho^2}\frac{1}{(1-\lambda
r^2)^2} \,
\label{qlam} \\
Q^{ij}&=&-\frac{8M^2\lambda a_i a_j (1+\lambda a_{(i}^2)(r^2+a_{j)} ^2)}{\rho^2(1-\lambda r^2)^2
(V-2M)(r^2+a_i^2)(r^2+a_j ^2)} \nonumber \\
&-&\frac{2M}{\rho^2}\frac{a_ia_j}{(r^2+a_i ^2)(r^2+a_j ^2)}
\nonumber
\\ &-& \frac{2M\lambda a_i a_j}{\rho^2(1-\lambda r ^2)} \left[
\frac{1}{(r^2 + a^2 _i)}+\frac{1}{(r^2 + a^2 _j )}\right] \nonumber
\\
 &+& \frac{4M^2 a_i a_j[(1+\lambda a_i ^2)+(1+\lambda
a_j^2)]}{\rho^2(1-\lambda r^2)^2 (V-2M)(r^2+a_i^2)(r^2+a_j^2)} \,.
 \label{qij}
\end{eqnarray}
Finally one more useful quantity is the determinant of the metric:
\bea\label{detg}
\sqrt{-g} = \frac{r\rho^2 \sin\theta \cos\theta}{\Xi_a \Xi_b}.
\eea
We should note that to compute these quantities it is easiest to use
the Kerr-Schild form of the metric and then perform a coordinate
transformation.

\section{The Hamilton-Jacobi equation}
The Hamilton-Jacobi equation for the problem at hand is
\begin{equation}
\frac{\partial S}{\partial l} + \frac{1}{2}g^{\mu\nu} \frac{\partial
S}{\partial x^{\mu}}\frac{\partial S}{\partial x^{\nu}} =0
\end{equation}
where $S$ is Hamilton's principal function. Recall it is a type II
generating function for a canonical transformation $(x^{\mu},
p_{\nu}) \to ( X^{\mu}, P_{\nu} )$ which implies $p_{\mu} = \partial
S / \partial x^{\mu}$ and $X^{\mu} = \partial S /
\partial P_{\mu}$.  Due to the presence of the Killing vectors we know that:
\begin{equation}
S = \frac{1}{2} m^2 l -E\tau + \sum_{i=1}^2 L_i \varphi_i +
F(r,\theta)
\end{equation}
and we will show that the problem is completely separable so
$F(r,\theta)= S_r(r)+S_{\theta}(\theta)$. The proof is as follows.
It is apparent that the inverse metric is largely composed of terms
which are of the form $f(r)/\rho^2$. Thus one is led to multiplying
the HJ equation by $\rho^2$ in order to achieve separability. The
only non-trivial terms are the first one in the function $Q$ and the
first term in $g^{\varphi_i \varphi_j}$. However, simple algebra
shows that
\bea \frac{\rho^2}{W(1-\lambda r^2)} = - \frac{\Xi_a
\Xi_b}{\lambda \Delta_{\theta}} + \frac{\Xi_a \Xi_b}{\lambda
(1-\lambda r^2)}
\eea
which takes care of the first term in $Q$. The identity
\bea \sum_{i=1}^2 \frac{L_i^2 \Xi_i}{\mu_i^2(r^2+a_i^2)}
\rho^2 &&= \sum_{i=1}^2 \frac{L_i^2 \Xi_i}{\mu_i^2} +
\frac{L_1^2 \Xi_a (b^2-a^2)}{r^2+a^2} \nonumber \\
&&+\frac{L_2^2 \Xi_b (a^2-b^2)}{r^2+b^2} \eea
takes care of the first term in $g^{\varphi_i \varphi_j}$. Thus, as promised,
multiplying the HJ equation by $\rho^2$ renders it separable. The
$\theta$ dependent part of the separated HJ equation reads:
\bea
\label{thetaHJ}
&&m^2(a^2\cos^2\theta +b^2 \sin^2\theta ) +\left(
\frac{\Xi_a L_1^2}{\sin^2\theta}+\frac{\Xi_b L_2^2}{\cos^2\theta} \right)
\nonumber \\&&+ \frac{\Xi_a \Xi_b}{\lambda \Delta_{\theta}} ( E- \lambda a_i L_i)^2 +
\Delta_{\theta} \left( \frac{\partial S_{\theta}}{\partial \theta}
\right)^2 = K. \eea The $r$ dependent part is rather more
complicated:
\bea \label{requ} &&(V-2M) \left( \frac{\partial S_r}{\partial r}
\right)^2  + \widetilde{V}(r;E,L_i,m) =-K,
\eea
where we have defined the ``effective'' potential
\bea
&&\widetilde{V}(r;E,L_i,m)= m^2 r^2 \nonumber \\ \nonumber &&- (E-\lambda a_i L_i)^2
\left( \frac{\Xi_a \Xi_b}{\lambda(1-\lambda
  r^2)} +\frac{2M}{(1-\lambda r^2)^2} \right) \\
&& + \rho^2 Q^{ij}L_iL_j -E^2\frac{4M^2}{(1-\lambda r^2)^2(V-2M)}
\nonumber \\
&& +\frac{8M^2 E L_i a_i(1+\lambda a_i
^2)}{(1-\lambda r^2)^2 (V-2M)(r^2+a_i ^2)} \nonumber \\
&& +4M E\frac{L_i a_i}{(1-\lambda
r^2)(r^2+a_i ^2)} \nonumber \\
&&+\frac{4M^2 L_i L_j a_i a_j (1+\lambda a_i^2) (1+\lambda a_j
^2)}{(1-\lambda r^2)^2(V-2M)(r^2+a_i ^2)(r^2 +a_j ^2)} \nonumber \\
&&\frac{L_1^2 \Xi_a (b^2-a^2)}{r^2+a^2}
+\frac{L_2^2 \Xi_b (a^2-b^2)}{r^2+b^2},
\eea
and $K$ is the separation constant.
Thus we have reduced the problem of solving for Hamilton's principal
function $S$ to quadratures.
Note that as $r \to \infty$ the potential $\widetilde{V} \sim m^2
r^2$ and $V \sim -\lambda r^4$; upon inspection of (\ref{requ}) this shows that when $\lambda <0$ (AdS) only bound orbits are possible.

We see that there exists an extra constant of motion $K$, as a
consequence of the separability of the HJ equation. This is due to
the presence of a Killing tensor which may be read off most easily
from (\ref{thetaHJ}), using $K=K^{\mu\nu}p_{\mu} p_{\nu}$ and
$g^{\mu\nu}p_{\mu}p_{\nu} = -m^2$ to give: \bea
K^{\mu\nu} &=& -g^{\mu\nu} ( a^2\cos^2\theta + b^2 \sin^2\theta) \nonumber \\
&+& \frac{\Xi_a \Xi_b}{\lambda \Delta_{\theta}}( \delta^{\mu}_{\tau}
\delta^{\nu}_{\tau} + 2\lambda \delta^{(\mu}_{\tau} \delta^{\nu
)}_{\varphi_i} a_i + \lambda^2 a_ia_j \delta^{( \mu}_{\varphi_i}
\delta^{\nu )}_{\varphi_j} ) \nonumber \\&+& \sum_{i=1}^2
\frac{\Xi_i}{\mu_i^2} \delta^{\mu}_{\varphi_i}
\delta^{\nu}_{\varphi_i} + \Delta_{\theta} \delta^{\mu}_{\theta}
\delta^{\nu}_{\theta}. \eea One may be concerned by the fact that
the $\lambda \to 0$ limit of this tensor does not exist, and thus
does not coincide with the Killing tensor of the five dimensional
Myers-Perry black hole found in~\cite{Frolov2}. However, one must
remember that one is free to add symmetrized outer products of
Killing vectors to a Killing tensor to give another Killing tensor.
In fact one can easily verify that the Killing tensor \bea
\widetilde{K}^{\mu\nu} = K^{\mu\nu} - \left( a^2+
b^2+\frac{1}{\lambda} \right) \delta^{\mu}_{\tau}
\delta^{\nu}_{\tau} - 2a_i \delta^{(\mu}_{\tau}
\delta^{\nu)}_{\varphi_i} \eea reduces to the correct flat space
limit. We should note that this is an \emph{irreducible} Killing
tensor. It is reasonable to ask whether, as in the four dimensional
case, the Killing tensor above can be decomposed into the square of a
Yano-Killing two-form $Y_{\mu \nu}$ as $K_{\mu \nu} = Y_{\mu
  \gamma}Y^{\gamma}_{\phantom{\gamma} \nu}$. It seems unlikely this is the
case, following the arguments in the case with one vanishing angular
momentum parameter~\cite{GLP2}.

For completeness we now give the general solution to geodesic motion
in the Kerr-(A)dS spacetime which can be easily deduced from the
generating function $S$ by differentiating with respect to $K, m^2, E, L_i$ respectively:
\bea
&&\int d\theta \frac{1}{\Delta_{\theta} S_{\theta}'(\theta)} = \int dr
  \frac{1}{(V-2M) S_r'(r)} \nonumber \\
l &=& \int d\theta \frac{a^2\cos^2\theta +b^2\sin^2\theta}{\Delta_{\theta}
  S_{\theta}'(\theta)} + \int dr \frac{r^2}{(V-2M)S_r'(r)} \nonumber \\
\tau &=& -(E-\lambda a_iL_i)\int d\theta \frac{\Xi_a\Xi_b}{\lambda
  \Delta_{\theta}^2  S_{\theta}'(\theta)} \nonumber \\ &-&\int dr \frac{ \partial_E
  \widetilde{V}(r;E,L_i,m)}{2(V-2M) S'_r(r)} \nonumber \\
\varphi_i &=& \int d \theta \frac{ 1
}{\Delta_{\theta}S_{\theta}'(\theta)} \left( \frac{\Xi_i L_i}{\mu_i^2}
  -\frac{\Xi_a \Xi_b a_i(E-\lambda a_jL_j)}{\Delta_{\theta}} \right) \nonumber \\
  &+& \int dr \frac{ \partial_{L_i}
  \widetilde{V}(r;E,L_i,m)}{2(V-2M) S'_r(r)}.
\eea
Finally we should remark that the phase space functions
$H,K,p_{\varphi_i},p_{\tau}$ are in involution under the Poisson
bracket, thus proving Liouville integrability.

\section{The Klein-Gordon equation}
Now we investigate the separability of the KG equation
\bea
\frac{1}{\sqrt{-g}} \partial_{\mu} ( \sqrt{-g} g^{\mu\nu}
\partial_{\nu} \Phi) = m^2 \Phi,
\eea describing a massive spinless field in the Kerr-(A)dS
background. Once again separability relies crucially on the fact that the functions
$\rho^{2}g^{\mu \nu}$ are separable. Using the
expressions~(\ref{ginv}) and~(\ref{detg}) one may express the KG
equation as
\bea\label{KG}
&&\frac{1}{r}\partial_{r}(r(V-2M)\partial_{r}\Phi) +
\frac{\partial_{\theta}(\Delta_{\theta}\sin{\theta}\cos{\theta}\partial_{\theta}\Phi)}{\sin{\theta}\cos{\theta}}
\nonumber \\
&&+ \rho^{2}g^{\tau \tau}\partial^{2}_{\tau}\Phi + 2\rho^{2}g^{\tau
  \varphi_{i}}\partial_{\tau}\partial_{\varphi_{i}}\Phi +
\rho^{2}g^{\varphi_{i}\varphi_{j}}\partial_{\varphi_{i}
  \varphi_{j}}\Phi \nonumber \\ && \qquad \qquad = m^{2} \rho^{2} \Phi.
\eea The obvious separation of variable ansatz gives
$\Phi~=~e^{-i\omega\tau}e^{im_{j}\varphi_{j}}R(r)\Theta(\theta)$ where
$m_1,m_2 \in \mathbb{Z}$. We
are then left with non-trivial equations for the functions $R(r)$
and $\Theta(\theta)$: \bea\label{Thetaeq}
&&\frac{ \frac{d}{d\theta} \left(
  \Delta_{\theta}\sin{\theta}\cos{\theta}\frac{d}{d\theta} \Theta
  \right)}{\Theta\sin{\theta}\cos{\theta}} \\
&&- \left(\frac{m_{1}^{2}\Xi_a}{\sin^{2}\theta} +
\frac{m_{2}^2\Xi_b}{\cos^{2}\theta}\right)-
\frac{\Xi_{a}\Xi_{b}}{\lambda\Delta_{\theta}} (\omega -\lambda
m_{i}a_{i})^2  \nonumber  \\&&- \nonumber m^2(a^{2} \cos^{2}\theta +
b^{2} \sin^{2}\theta)  = k, \eea where $k$ is the separation
constant. Now let us briefly analyse this equation. Firstly, we
change variables to $z=\sin^2\theta$ which gives: \bea &&\frac{d^2
\Theta}{dz^2} + \left( \frac{1}{z} + \frac{1}{z-1}+ \frac{1}{z-d}
\right) \frac{d
    \Theta}{dz} \nonumber \\
&&- \left[ \frac{1}{4 \Delta_z z(1-z)} \left( \frac{m_1^2\Xi_a}{z}+
\frac{m_2^2\Xi_b}{1-z} \right) + \frac{\Xi_a \Xi_b(\omega-\lambda
a_i m_i)^2}{4\lambda
  \Delta_z^2 z(1-z)} \right. \nonumber \\
&&\left. + \frac{m^2}{4\lambda z(1-z)}+ \frac{k'}{4\Delta_z
  z(1-z)} \right] \Theta =0,
\eea where we have $k'=k-\frac{m^2}{\lambda}$, $d=-
\frac{\Xi_a}{\lambda(b^2-a^2)}$ and $\Delta_z~=~\Xi_a+~\lambda
(b^2-~a^2)z$. We immediately recognise this as a second order
Fuchsian equation. It is easily verified that it has four regular
singular points located at $z=0,1,d,\infty$. Therefore by a
transformation of the form $\Theta(z)~=~z^{A}(z-~1)^B(z-d)^C y(z)$,
one may show that $y(z)$ satisfies Heun's equation for suitable
choice of $A,B,C$. A welcome simplification occurs in the degenerate
case $a=b$. Then it is easy to see that the resulting equation for
$\Theta(z)$ only has three regular singular points located at
$z=0,1,\infty$. Thus the solutions are immediately expressible in
terms of hypergeometric functions. The only solution which is
regular at $\theta=0$ is: \bea &&\Theta(z) = N z^{\alpha_1}
(1-z)^{\beta_1}  {}_2 F_1 (\alpha,\beta,\gamma;z), \\ \nonumber
&&\alpha_1 = \frac{|m_1|}{2}, \qquad \beta_1 = \frac{|m_2|}{2}, \\
\nonumber &&\alpha= \alpha_1+\beta_1+\gamma_1, \qquad \beta=
\alpha_1+\beta_1 +1-\gamma_1,\\ \nonumber && \gamma= 1+2\alpha_1,
\eea where $(\alpha_1,-\alpha_1)$, $ (\beta_1, -\beta_1)$,
$(\gamma_1, 1-\gamma_1)$ are the indices of the equation for
$\Theta$ at $0,1,\infty$ respectively. The indices at infinity are
solutions to the quadratic equation: \bea x^2 -x +
\frac{(\omega-\lambda a_i m_i)^2}{4\lambda} + \frac{m^2}{4\lambda}
+\frac{k'}{4 \Xi_a} =0. \eea Requiring that the solution is also
regular at $\theta=\pi/2$ implies that $\beta=-n$ where $n \in
\mathbb{N}$, which can be thought of as a quantization condition of
the separation constant $k$. This implies one can write the solution
in terms of Jacobi polynomials: \bea \Theta(z)= N' z^{\alpha_1}
(1-z)^{\beta_1} P_n^{(2\alpha_1,\,
    \alpha-\gamma-n)}(1-2z).
\eea

For completeness we give the radial equation (valid for general rotation parameters) which may be expressed in the form:
\bea\label{KGradial}
&-&\frac{1}{r} \frac{d}{dr} \left( r(V-2M) \frac{d}{dr}R(r) \right)
\nonumber \\ && \qquad +\widetilde{V}(r;\omega,m_i,m)R(r)= kR(r).
\eea
This is also a second order Fuchsian equation. We will not concern
ourselves with its analysis here.


\section{Comments}
An interesting question is whether these integrability results we have
found generalise to $D>5$ for general rotation parameters. As remarked
earlier for $D>5$ the best result to date is when all rotation
parameters are set equal, or when only one is non-zero. The situation
for the asymptotically flat black holes of Myers and Perry is a little
better; a slightly stronger result has been obtained where
integrability has been proved when the rotation parameters $a_i$ take on
at most two different values~\cite{Palmer,Page2}. This gives rise to an
irreducible Killing tensor. The difficulty in tackling
the higher dimensional cases rests on the fact that the metric
contains cross terms of the form $d\mu_i d\mu_j$. A consequence of
separability of the $D>5$ black holes would be the presence of more
Killing tensors. For suppose the HJ equation was separable for the
general Myers-Perry black hole. Then we know that the number of
Killing vectors is $1+ [(D-1)/2]$. Together with the geodesic
constraint this implies that to achieve integrability, which requires
$D$ constants of motion, one would need $D-2-[(D-1)/2]$ Killing
tensors. Thus for $D>5$ the number of Killing tensors would be
greater than one.

Another generalisation of~\cite{Frolov2} would be to investigate the
existence of Killing tensors in the vacuum black ring~\cite{Emparan} which has the
same isometry group as the five dimensional Myers-Perry black
hole. However the coordinates in which it is usually written do not
appear to be adapted to separability. \\
\indent Finally, we note that for $\lambda <0$ the black hole considered here
is asymptotically AdS. These black holes are known to be dual to a
thermal conformal field theory (CFT) on $S^3$~\cite{Berman}. It is well known that isometries in the bulk
theory give rise to conserved charges on the dual CFT living on the
boundary. It would be interesting to see whether the existence of Killing
tensors may correspond to some analogous quantity in the dual
theory.

\begin{acknowledgements}
We would like to thank Gary Gibbons for useful comments and reading
through the manuscript. HKK would like to thank St. John's College,
Cambridge, for financial support.
\end{acknowledgements}

\end{document}